\documentclass[12pt]{JHEP}
\usepackage{graphics}
\usepackage{cite}
\newcommand{\beq}{\begin{equation}}
\newcommand{\eeq}{\end{equation}}
\newcommand{\bea}{\begin{eqnarray}}
\newcommand{\eea}{\end{eqnarray}}
\renewcommand{\d}{\delta}

\renewcommand{\b}{\beta}

\newcommand{\m}{\mu}

\newcommand{\oh}{\frac{1}{2}}

\newcommand{\dg}{\dagger}
\newcommand{\non}{\nonumber}

\newcommand{\rf}[1]{(\ref{#1})}
\newcommand{\ra}{\rightarrow}

\title{Remarks on the Gribov Problem in Direct Maximal Center Gauge }

\author{Manfried Faber \\ Inst. f\"ur Kernphysik, 
Technische Universit\"at Wien, 
A-1040 Vienna, Austria \\
E-mail: \email{faber@kph.tuwien.ac.at}}

\author{Jeff Greensite \\ Physics and Astronomy Department, San Francisco
State University, San Francisco, CA 94117 USA 
\\ Theory Group, Lawrence 
Berkeley National Laboratory, Berkeley, CA 94720 USA \\
E-mail: \email{greensit@quark.sfsu.edu,JPGreensite@lbl.gov}}

\author{{\v S}tefan Olejn\'{\i}k \\ Institute of Physics, Slovak Academy 
of Sciences, SK-842 28 Bratislava, Slovakia \\
E-mail: \email{fyziolej@savba.sk}}

\abstract{ We review the equivalence of maximal center gauge fixing to
the problem of finding the best fit, to a given lattice gauge field,
by a thin vortex configuration.  This fit is necessarily worst at the
location of P-plaquettes.  We then compare the fits achieved in Gribov
copies generated by (i) over-relaxation; (ii) over-relaxation after
Landau gauge preconditioning; and (iii) simulated annealing.
Simulated annealing yields the best fit if all links on the lattice
are included, but the situation changes if we consider only the
lattice volume exterior to P-plaquettes.  In this exterior region, the
fit is best for Gribov copies generated by over-relaxation, and worst
for Gribov copies generated after Landau gauge preconditioning. 
The two fitting criteria (including or not including the P-plaquettes)
yield string tensions differing by $-34$\% to $+20$\% respectively,
relative to the full string tension.  Our usual procedure (``quenched
minimization'') seems to be a compromise between these criteria, and
yields string tensions at an intermediate value close to the full
string tension.}


\begin{document}

\section{Introduction}

   Direct maximal center gauge is the common name for lattice Landau
gauge in the adjoint representation.  For the SU(2) group, adjoint links
$U_A$ in this gauge satisfy the condition
\beq
      \mbox{Tr} \sum_\m L_i 
         \Bigl(U_{A\m}(x) - U_{A\m}(x-\hat{\m})\Bigr) = 0
\label{dmc1}
\eeq
at every point (the $\{L_i\}$ are the SU(2) group generators in
the adjoint representation) leaving a residual $Z_2$    
symmetry.  This gauge is used in studies of confinement, in particular
to locate center vortices in thermalized lattice configurations.

   As in ordinary lattice Landau gauge, there are many points on the
gauge orbit, not related by any residual symmetry, which satisfy the
local gauge condition \rf{dmc1}.  These configurations are known as
Gribov copies.  Difficulties associated with Gribov copies in 
maximal center gauge have been noted in recent work by Bornyakov et al.\
\cite{BKP,BKPV}, and previously by Kov\'{a}cs and Tomboulis 
\cite{KT}.  In this article we discuss some issues that we believe
are relevant to the problem of generating a set of Gribov copies
in maximal center gauge, and to choosing the ``best''
Gribov copy among the set.  Before proceeding, we should note that
the Gribov problem can be avoided altogether by using the Laplacian center
gauge \cite{Pepe}.  On the other hand, the Laplacian version does have
certain unattractive features, notably the lack of scaling of the
P-vortex density \cite{Rein} and the lack of ``precocious linearity'' 
\cite{mog}. While these features of Laplacian center gauge 
are by no means fatal, we think it worthwhile to further 
explore the Gribov copy issue in the older proposal of maximal 
center gauge.

\section{Gauge Fixing as a ``Best Fit'' Procedure}

   Gauge fixing always has a flavor of arbitrariness, and claims for a
privileged gauge are typically regarded with suspicion. To understand
the rationale for maximal center gauge, it is essential to understand
its function as a fit of a given lattice configuration by a singular
pure gauge (thin vortex) field.  This insight, which we will now
elaborate, is due to Engelhardt and Reinhardt in ref.\ \cite{ER}.

   Imagine running a Monte Carlo simulation of lattice SU(2) gauge theory
at high $\b$, and printing out the values of link variables $U_\m(x)$
in some thermalized
configuration.  At a glance, these values would look like random numbers,
but of course this impression 
is deceptive.  Locally the link variables are only small
fluctuations around a pure gauge configuration.  Suppose we then ask for
the pure gauge configuration $g(x) g^\dg(x+\hat{\m})$ which is closest, in
lattice configuration space, to the given lattice gauge field $U_\m(x)$.
Using the standard metric on the SU(2) group manifold, this is equivalent
to asking for the gauge transformation $g(x)$ which minimizes the
square distance in configuration space 
\bea
      d^2 &=& \sum_{x,\m} \mbox{Tr}\Bigl[(U_\m(x)-g(x) g^\dg(x+\hat{\m}))
                        (U_\m(x)-g(x) g^\dg(x+\hat{\m}))^\dg \Bigr]
\non \\
          &=& 2 \sum_{x,\m} \Bigl(\mbox{Tr}[I_2] - 
                          \mbox{Tr}[g^\dg(x)U_\m(x)g(x+\hat{\m})]\Bigr)
\eea
This quantity is minimized by finding the gauge-transformed configuration
\beq
         {}^gU_\m(x) \equiv g^\dg(x)U_\m(x)g(x+\hat{\m})
\eeq
which maximizes 
\beq
      R_{lan} = \sum_{x,\m} \mbox{Tr}[{}^gU_\m(x)]
\eeq
From this we conclude that the problem of finding the pure-gauge configuration
closest to a given lattice gauge field is completely equivalent to the problem
of fixing that lattice gauge field to the Landau gauge.

   We next generalize this idea slightly, and allow for $Z_2$ dislocations
in the gauge transformation.  This means fitting the lattice configuration
by a slightly more general form
\beq
        U^{vor}_\m(x) = g(x) Z_\m(x) g^\dg(x+\hat{\m})
\label{Uvor}
\eeq
where $Z_\m(x) = \pm 1$.  This is a thin center vortex configuration,
generated by a singular gauge transformation.  Note that $U_\m^{vor}$
becomes a continuous pure gauge in the adjoint representation, which
is blind to the $Z_\m(x)$ factor.  This motivates a two-step fitting
procedure:  First, we determine $g(x)$ up to a residual $Z_2$ transformation,
by minimizing the square distance $d^2_A$ in configuration space 
between $U_\m(x)$
and $U^{vor}_\m(x)$ in the adjoint representation:
\bea
      d_A^2 &=& \sum_{x,\m} 
           \mbox{Tr}\Bigl[(U_{A\m}(x)-g_A(x) g_A^\dg(x+\hat{\m}))
                        (U_{A\m}(x)-g_A(x) g_A^\dg(x+\hat{\m}))^\dg \Bigr]
\non \\
          &=& 2 \sum_{x,\m} \Bigl(\mbox{Tr}[I_3] - 
               \mbox{Tr}[g_A^\dg(x)U_{A\m}(x) g_A(x+\hat{\m})]\Bigr)
\non \\
    &=& 2 \sum_{x,\m} \Bigl( 4 - 
               \Bigl|\mbox{Tr}[g^\dg(x)U_{\m}(x) g(x+\hat{\m})]\Bigr|^2\Bigr)
\eea
The distance function in this case is determined by the metric on the
$SU(2)/Z_2$ group manifold.  Minimizing $d_A^2$ is
equivalent to fixing to direct maximal center gauge, which seeks the largest
value of
\beq
      R_{dmc} = \sum_{x,\m} \Bigl|\mbox{Tr}[{}^gU_\m(x)]\Bigr|^2
\eeq
The condition that $R$ is stationary leads to the adjoint Landau gauge
condition \rf{dmc1}.
Having determined $g(x)$, we then find $Z_\m(x)$ at each link by
minimizing
\beq
      l^2_{\m}(x) = \mbox{Tr}\Bigl[(U_\m(x)-g(x) Z_\m(x)g^\dg(x+\hat{\m}))
                        (U_\m(x)-g(x)Z_\m(x)g^\dg(x+\hat{\m}))^\dg \Bigr]      
\eeq
which is easily seen to require
\beq
        Z_\m(x) = \mbox{signTr}[{}^gU_\m(x)]
\label{cp}
\eeq
This is the center projection prescription.

   In this way we have shown that maximal center gauge, together with
the rule \rf{cp} for center projection, is equivalent to finding a
``best fit'' of a given lattice configuration $U_\m(x)$ by a thin vortex
configuration $U^{vor}_\m(x)$.  While this (rather trivial) derivation may
be novel, the essential point 
has been made previously in ref.\ \cite{ER}.

\subsection{Center Dominance}

   The gauge-transformed lattice configuration in maximal center gauge 
can be written
\beq
   {}^gU_\m(x) = Z_\m(x) e^{iA_\m(x)} ~~~~~~~ (\mbox{Tr}e^{iA_\m(x)} \ge 0)
\label{U}
\eeq
with the original lattice configuration 
\beq
       U_\m(x) = g(x) Z_\m(x) e^{iA_\m(x)} g^\dg(x+\hat{\m})
\label{gU}
\eeq
where $g(x),Z_\m(x)$ are determined by the procedure just described.
Our claim, elaborated in refs.\ \cite{mog,Jan98}, is that the confining
properties of $U_\m(x)$ are entirely encoded in the $Z_2$ link variables
$Z_\m(x)$; the variables $A_\m(x)$, which are typically of order $1/\b$, 
are responsible
for short range effects such as the Coulomb force law.  
Two necessary, but not sufficient, conditions for this claim to hold true
are as follows:
\begin{enumerate}
\item Set $A_\m(x)=0$ in \rf{gU}, which sends $U_\m(x)\ra U^{vor}_\m(x)$.
The asymptotic string tension in the $U^{vor}$ configuration should
match the asymptotic tension of the full configuration, i.e.
``center-projected'' Wilson loops
\beq
       Z(C) = \oh \mbox{Tr}[U^{vor}(C)]
\eeq 
have the same string tension as full Wilson loops (where $Z(C)$ denotes
a product of center-projected links around loop $C$).  
This property is known as ``Center Dominance.''
\item  Dropping the $Z_\m(x)$ variables in \rf{gU}, which can be 
interpreted as removing center vortices from the system, the Wilson loops
constructed from the remaining degrees of freedom
\bea
       \tilde{W}(C) &\equiv& \mbox{Tr}[\prod_{links\in C} e^{iA_\m}]
\non \\
            &=& Z(C) \mbox{Tr}[U(C)]
\eea
should have zero string tension.  This test was carried out 
by de Forcrand and D'Elia in ref.\ \cite{dFE}.
\end{enumerate}
 
  The reason that these properties are not sufficient for our purposes
is that confining fluctuations might be hidden in the $Z_\m(x)$
variables along with a lot of short-range physics.  If that is the
case, then we might not gain much from the truncation of $U_\m$ to
$Z_\m$.  For example, even a naive projection $Z_\m(x) =
\mbox{signTr}[U_\m(x)]$ of the original configuration (with no gauge
fixing) has the center dominance property, as is easily verified by
simple group-theoretic arguments \cite{Ogilvie,mog}.  But this naive
projection also reproduces $-$ exactly! $-$ the Coulomb potential at
short distance scales, which is certainly due to gaussian field
fluctuations rather than a vortex mechanism.  Our objective is to
strip away such short-range effects, and to isolate as far as possible
the degrees of freedom which are solely responsible for infrared
physics.

\subsection{Where Fitting Fails}

   The square deviation at each link, away from a pure gauge in the
adjoint representation, is proportional to
\bea
     \d^2_{\m}(x) &=& {1\over 8} 
        \mbox{Tr}\Bigl[(U_{A\m}(x)-g_A(x) g_A^\dg(x+\hat{\m}))
                        (U_{A\m}(x)-g_A(x) g_A^\dg(x+\hat{\m}))^\dg \Bigr]
\non \\
      &=& 1 - {1\over 4} \Bigl(\mbox{Tr}[{}^gU_\m(x)]\Bigr)^2
\eea
This quantity 
is generally small, of order $1/\b$, at a local minimum of $d^2_A$.
The exception is for some links (at least one) belonging
to each P-plaquette.  For these links we must have a very poor fit, 
with $\d_\m^2(x) \sim O(1)$ regardless of $\b$.  We recall that a 
plaquette $p$ is a P-plaquette iff $Z(p)=-1$, and that  P-plaquettes belong
to P-vortices.  
   
   The fit to a thin vortex has to be bad at P-plaquettes 
for the following reason:
At large $\b$, we generally have (even at P-plaquettes)
\beq
      \oh \mbox{Tr}[U(p)] = 1 - O({1\over \b})
\label{one}
\eeq
On the other hand
\beq
 \mbox{Tr}[U(p)] = Z(p) \mbox{Tr}[\prod_{links \in p}e^{iA_\m(x)}]
\label{two}
\eeq
But if $Z(p)=-1$, 
equations \rf{one} and \rf{two} imply that $A_\m(x)$ is $O(1)$
rather than $O(1/\b)$, on one or more links in the plaquette $p$.
The magnitude of $A_\m(x)$ is a measure of the goodness-of-fit,
and in fact $\d_\m^2$ depends only on this variable.  A perfect fit on a 
link corresponds to $A_\m=0$, while $A_\m \sim O(1)$ is a very bad
fit.

   In ordinary Landau gauge, $\d^2_\m(x)$ can in principle
be small on every link.  In contrast, in maximal center gauge, we
see that this property can only 
hold if there are no P-vortices found on the lattice.
   
\section{Quenched vs.\ Annealed Gribov Copies}

   There is no known method for finding the global minimum of the
distance function $d^2_A$, but two methods have been employed to
generate local minima, i.e.\ Gribov copies, satisfying the local
condition \rf{dmc1}.  These are the methods of simulated quenching,
and simulated annealing.  The terms derive from the fact that minimizing
$d^2_A$ is equivalent to finding the zero-temperature ground state
of an analog ``spin-glass'' system
\bea
      H &=& - \sum_{x,\m} 
        \mbox{Tr}\Bigl[g^\dg_A(x) U_{A\m}(x) g_A(x+\hat{\m})\Bigr]
\non \\
        &=& - \sum_{x,\m} \left( 
   \mbox{Tr}\Bigl[g^\dg(x) U_{\m}(x) g_(x+\hat{\m})\Bigr]\right)^2
        + \mbox{const}
\label{ss}
\eea
where $g_A(x)$ is the dynamical SO(3) group-valued ``spin'' variable,
and $U_A$ is a set of fixed, stochastic, nearest-neighbor couplings.

   An obvious approach to finding the zero-temperature ground state
is to cool the system, either gradually (``annealing'') or suddenly
(``quenching'').  Quenching corresponds to placing the system in contact
with a reservoir at zero temperature, and implies 
that only changes in $g(x)$ which lower the energy density 
can be accepted.  In practice, quenching
is implemented by the over-relaxation method, as described
in ref.\ \cite{Jan98}.  Simulated annealing is achieved by applying Metropolis
updates to the analog spin system, and gradually lowering the temperature 
variable from some initial value to zero \cite{BKP}.  Neither of these
methods obtains the true minimum of $H$; each generates a set
of Gribov copies in maximal center gauge.

   A variation of the quenching approach was studied by 
Kov\'{a}cs and Tomboulis \cite{KT}.  Instead of applying over-relaxation
to a random point on the gauge orbit of a thermalized configuration,
they first fixed the configuration to ordinary lattice Landau gauge,
and then applied over-relaxation to fix to maximal center gauge.  We will 
refer to the quenching procedure via over-relaxation described in
\cite{Jan98} as ``OR'', the 
Kov\'{a}cs-Tomboulis variation of this procedure as ``KT'', and simulated
annealing, studied in maximal center gauge by Bornyakov et al.\ \cite{BKP},
as ``SA''.

   In Table \ref{tab1} we display the average deviation $\d^2$ 
per link
\beq
       \d^2 = {1\over 4V} \sum_{x,\m} \d^2_\m(x)
\eeq
for Gribov copies generated by the KT, SA, and OR methods ($4V$ is the
number of links on the lattice).
These methods were applied to thermalized lattice SU(2) configurations 
generated at the $\b$ values and lattice sizes shown below.
Only one Gribov copy was generated, in each method, for each thermalized 
lattice; there was
no attempt to select the ``best'' copy out of a set.
We also display, in Fig.\ \ref{r}, the fractional deviation 
\beq
      f = {\d^2 - \d^2(OR) \over \d^2(OR)}
\eeq   
of $\d^2$ away from the OR result obtained in each method.
We see that, according to the criterion of minimizing the average
$\d^2$, simulated annealing gives the best fit to a thin vortex,
the KT method is a little worse, and OR is the worst of the three, as
already noted in ref.\ \cite{BKP}.

\TABLE[h]{
\centerline{
\begin{tabular}{|c|c|c|c|c|} \hline
  $\b$ & lattice size   & Kovacs-Tomboulis  & sim anneal & pure over-relax 
       \\ \hline
  $2.1$ & $10^4$  & $0.2797(3)$ & 0.2736(3) & 0.2805(4)  \\
  $2.2$ & $12^4$  & $0.2631(2)$ & 0.2591(2) & 0.2657(2)  \\
  $2.3$ & $16^4$  & $0.2440(1)$ & 0.2406(1) & 0.2467(1)  \\
  $2.4$ & $20^4$  & $0.2232(2)$ & 0.2206(2) & 0.2260(2)  \\  \hline 
\end{tabular} } 
\caption{Values for the average $\d^2$, for Gribov copies
generated by three different methods.}
\label{tab1}
}

\FIGURE[h]{
\centerline{\scalebox{1.0}{\includegraphics{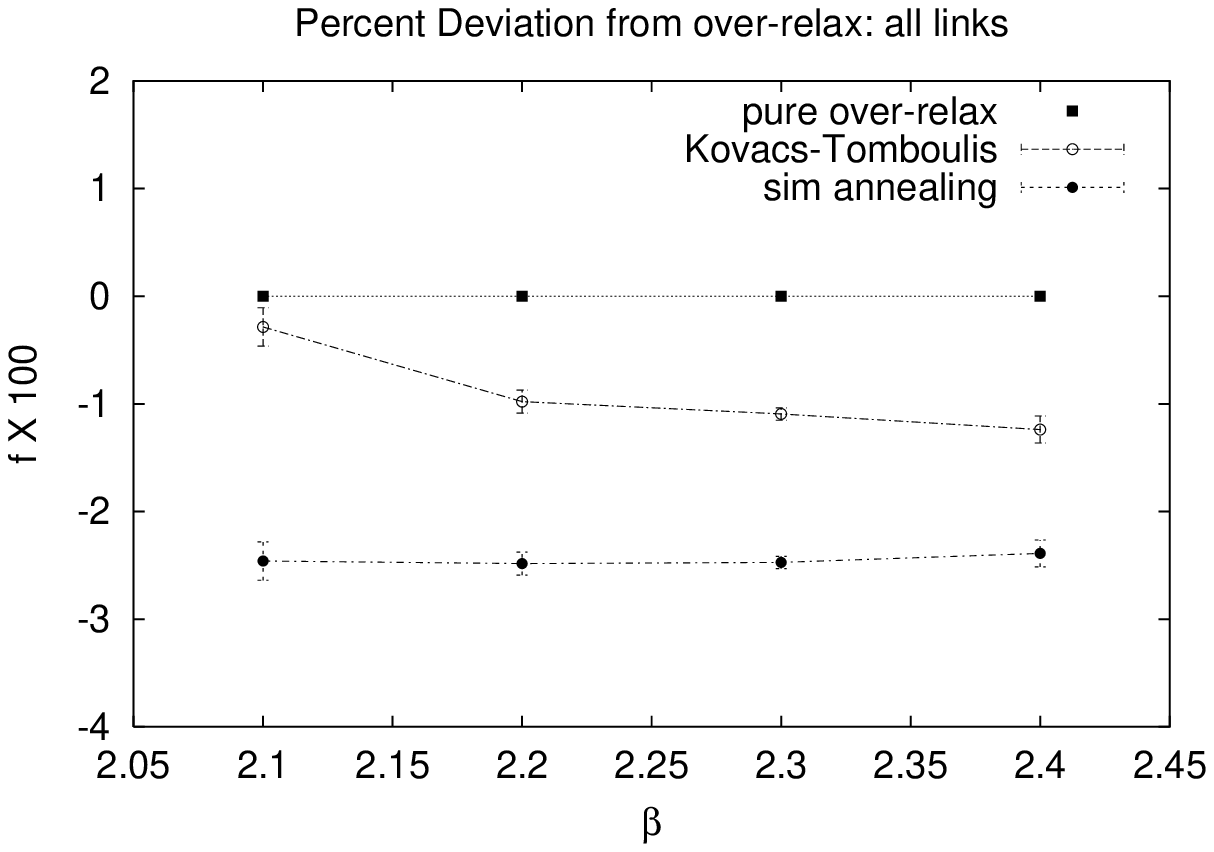}}}
\caption{Percent deviation ($f\times 100$) of $\d^2$ from the
over-relaxation value, for the simulated annealing and
Kov\'acs-Tomboulis methods.  Simulated annealing is best at minimizing
$\d^2$.}
\label{r}
}

   There is something a little peculiar about this ordering, however.
The OR Gribov copies are known to have reasonably good center dominance
properties \cite{Jan98,aborny}.
Gribov copies generated by the KT method result in a vanishing projected
string tension \cite{KT}, while the SA procedure yields projected
tensions at roughly 66\% of the full string tension 
\cite{BKP}.  One might have expected
that the projected string tension would be correlated directly with
$\d^2$.  Instead, what seems to happen is that the
projected string tension first drops to zero,  
as $\d^2$ falls below the OR result in KT copies, 
and then rises again to
66\% of the full string tension as $\d^2$ falls further in SA copies.  

   This raises the question of whether there is some other aspect of the fit to
a thin vortex which is better correlated with the projected string
tension.  We have already noted that the fit is necessarily very bad
at the location of P-plaquettes. This suggests just ignoring
the quality of fit at the P-plaquettes, where it is guaranteed to
be bad, and concentrating on the fit in the lattice volume exterior to
P-plaquettes.  We therefore calculate the average square deviation
in the exterior region
\beq
     \d^2_{ext} = {1\over N_{ext}}\sum_{ext} \d^2_\m(x) 
\eeq
where $N_{ext}$ is the number of all 
links not belonging to P-vortices, and the sum runs over the set of these
external links.  The quantity $\d^2_{ext}$ is
a measure of the quality of fit to a thin vortex in this exterior region.

   Table \ref{tab2} displays the values of $\d^2_{ext}$, and Fig.\ \ref{rq}
the corresponding fractional deviation 
\beq
      f_{ext} = {\d^2_{ext} - \d^2_{ext}(OR) \over \d^2_{ext}(OR)}
\eeq   
in the external region.  Here the order is quite different than in
Table \ref{tab1} and Fig.\ \ref{r}.  This time, the best fit is achieved by
Gribov copies generated by quenching (OR), and these copies also have
the largest projected string tension.  The SA fit is a little worse, with
the projected string tension $\approx 34\%$ lower than the full string
tension, while KT copies have by far
the worst fit, and also have negligible projected string tension.  From this
data, it appears that the projected string tension correlates much better
with the fit in the external region, based on $\d^2_{ext}$, 
than with the overall fit $\d^2$, which includes the P-vortex volume.

\TABLE[h]{
\centerline{
\begin{tabular}{|c|c|c|c|c|} \hline
  $\b$ & lattice size & Kovacs-Tomboulis  & sim anneal & pure over-relax 
       \\ \hline
  $2.1$ & $10^4$  & $0.2062(4)$ & 0.1951(3) & 0.1936(3)  \\
  $2.2$ & $12^4$  & $0.2081(4)$ & 0.1970(3) & 0.1944(3)  \\
  $2.3$ & $16^4$  & $0.2086(1)$ & 0.1975(2) & 0.1948(1)  \\
  $2.4$ & $20^4$  & $0.2043(1)$ & 0.1950(1) & 0.1927(1)  \\  \hline 
\end{tabular} } 
\caption{Values for $\d^2_{ext}$.}
\label{tab2}
}

\FIGURE[h]{
\centerline{\scalebox{1.0}{\includegraphics{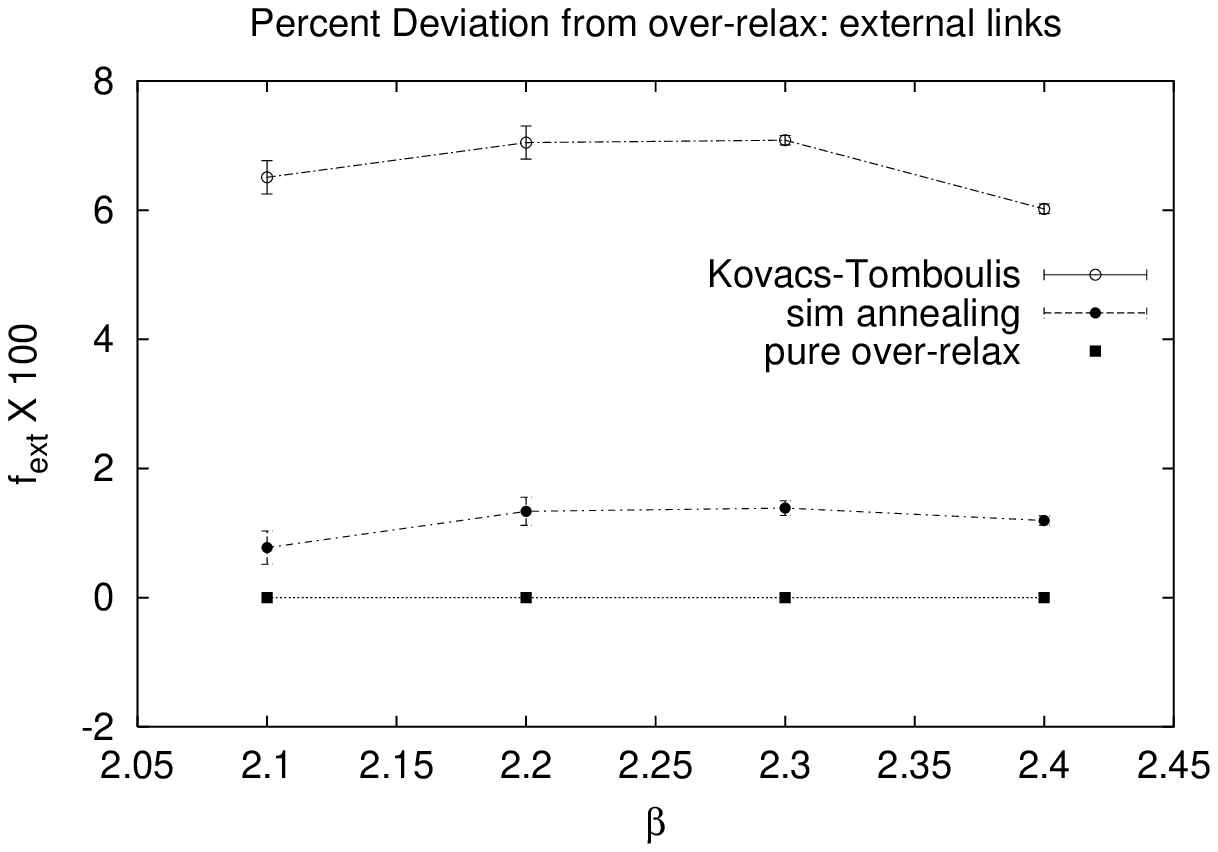}}}
\caption{Same as Fig. \ref{r}, but for the exterior region.  
Pure over-relaxation is best at minimizing $\d^2_{ext}$.}
\label{rq}
}

The results just obtained suggest selecting among Gribov copies
generated by quenching, on the basis of the smallest average
$\d^2_{ext}$.  But this procedure is also not a great success as regards
center dominance.  
We have found that choosing the lowest $\d^2_{ext}$ 
out of a set of quenched copies leads to projected string tensions
which are $\approx 20\%$ higher ($\b=2.4$) than the full string tension, 
as seen, e.g., from the projected Creutz ratio $\chi_{cp}(5,5)$ shown in
Fig.\ \ref{RQG}.

\FIGURE[h]{
\centerline{\scalebox{0.9}{
{\includegraphics{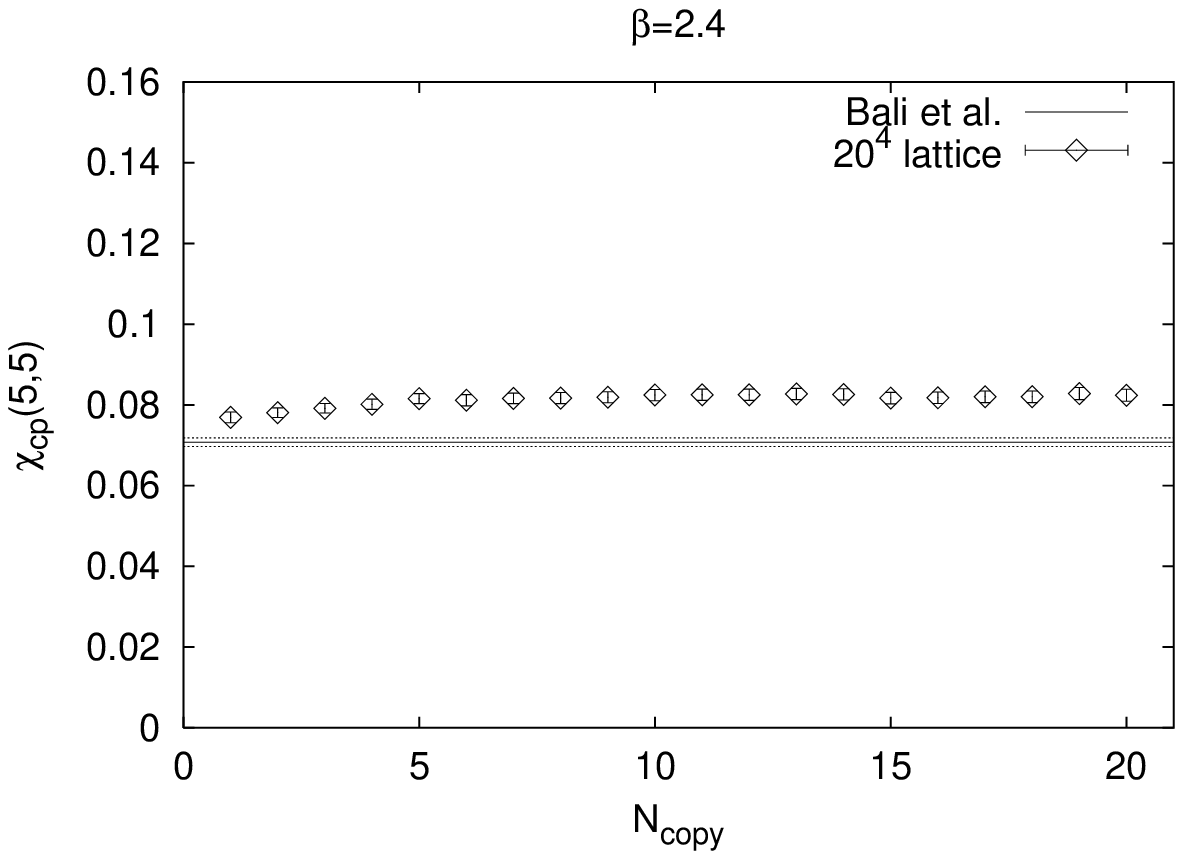}}}}
\caption{$\chi_{cp}(5,5)$ vs.\ the number $N_{copy}$ of Gribov copies
generated per configuration, at $\b=2.4$.  The Creutz ratio is evaluated in the
copy with the best exterior fit, minimizing $\d_{ext}^2$.  Solid and dotted
lines are the asymptotic string tension and errorbars, respectively, on the
unprojected lattice \cite{Bali}.}
\label{RQG}
}

\section{Quenched Minimization}

    If the criterion for the best gauge copy is the best average fit
over all links, including P-vortices, then the projected string
tension is about 34\% lower than the full string tension, as shown
by Bornyakov et al.\ in ref.\ \cite{BKP}.  On the other hand, if the
criterion for best fit is the fit to a thin vortex in the region
exterior to P-vortices, then the projected string tension comes out around
20\% too high ($\b=2.4$).  A case can be made for either fitting criterion, but
the method which we have used in the past now appears to be something of 
a compromise between the two.  We begin by quenching the analog spin
system \rf{ss}, starting from each of a set of random points on the
gauge orbit, to generate a set of Gribov copies satisfying \rf{dmc1}.
This procedure, as seen in the previous section, tends to emphasize
the exterior fit $\d^2_{ext}$ to a thin vortex.  Among the gauge copies
generated by quenching, we then select the copy with the best overall 
fit; i.e.\ the minimum value of $\d^2$.

   In the absence of a compelling reason to prefer the $\d^2$ or
$\d^2_{ext}$ criterion, the ``quenched minimum'' prescription
seems a priori as good as any.  It is, at least, a perfectly
well-defined procedure.  We have argued in the past, on the basis
of a number of empirical tests, that center projection applied to quenched
minima locate physical objects, namely the center vortices.  The results
of those tests have not changed any, and are worth summarizing once again:

\begin{enumerate}
\item  Let $W_n(C)$ denote the expectation
value of a Wilson loop constructed from unprojected links, evaluated
in the subensemble of configurations in which $n$ P-vortices, on the
projected lattice, pierce the minimal area
of loop $C$.  It is found that
\beq
     {W_n(C) \over W_0(C)} \ra (-1)^n
\eeq
as expected if P-vortices locate center vortices in the
unprojected configurations \cite{Jan98}.
\item If a vortex is inserted ``by hand'' (via a singular gauge 
transformation) into a thermalized lattice, then the set of P-vortices on the
projected lattice includes the inserted vortex \cite{Yamada}. 
\item Using information about P-vortex location to remove
center vortices from the unprojected configuration, one finds that
the confining and chiral symmetry breaking properties of the
configuration are also removed, and the topological charge goes
to zero \cite{dFE}.
\item The density of P-vortices, and therefore the density of
vortices identified on the unprojected lattice, scales correctly
according to the renormalization group, as first noted in \cite{Tubby1}
(see also the later results in \cite{Jan98,mog}).
\item Center dominance: The string tension of the projected lattice agrees 
fairly well with the asympotic string tension of the unprojected lattice 
\cite{Jan98,aborny}.
\item Precocious linearity:  Projected Creutz ratios $\chi_{cp}(I,I)$
vary only a little with $I$, and there is no Coulombic force at
small distances.  This indicates that the projected degrees of
freedom are not mixed up with short-range physics \cite{mog}.  
\end{enumerate}

   Despite these apparent successes, there is still a serious
objection that can be raised to the quenched minimization approach, in view
of the findings of Bornyakov et al.\ \cite{BKP}.  The problem is that
\emph{any} local minimum of $d^2_A$ can be reached by quenching, if the
starting configuration is within the ``basin of attraction'' of that
minimum on the gauge orbit.  
Thus, if we sample enough random configurations on the gauge
orbit by the OR approach, eventually minima obtained by the SA method
would be reached, and the projected string tension would drop well
below the full string tension.

   The answer to this objection is based on the fact that 
the projected string tension is found to converge rapidly,
with the number of Gribov copies generated, to a value
in good agreement with the full string tension.  Moreover, the convergence
improves as the lattice size increases \cite{aborny}.  The implication
is that the measure of SA copies must be negligible compared to the
measure of OR copies, at least at large volume.  Both
the volume and copy-number dependence of the projected Creutz ratio
$\chi_{cp}(5,5)$ at $\b=2.5$ are illustrated in Fig.\ \ref{MCG}, taken
from ref.\ \cite{aborny}.  Similar examples of other Creutz ratios, at various
$\b$ values, can also be found in that reference.  The indications
are that on an infinite lattice there is convergence to a result well above
the SA value.  The probability
of a random configuration evolving, under quenching, to an SA minimum
seems likely to go to zero in the infinite volume limit (although this
conjecture deserves further investigation).

\FIGURE[h]{
\centerline{\scalebox{1.0}{
{\includegraphics{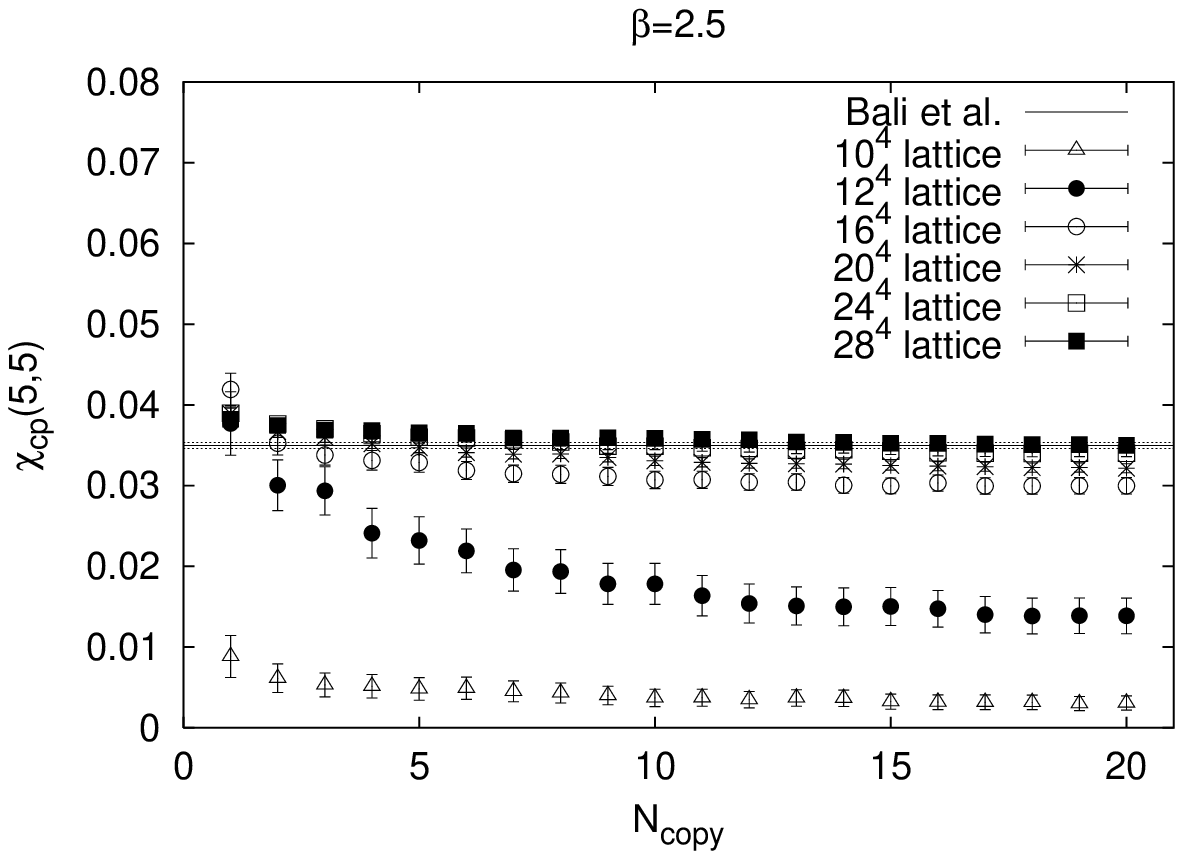}}}}
\caption{$\chi_{cp}(5,5)$ vs.\ the number $N_{copy}$ of Gribov copies
generated per configuration, obtained at various lattice volumes.  
The quenched minimization procedure is used.}
\label{MCG}
}

   A last point which we would like to make, in connection with
quenched Gribov copies, is that the projected lattices of different
copies seem to be quite well correlated at distances beyond one fermi,
which is the width of a center vortex.\footnote{The width of center
vortices can be determined in three different ways: from the falloff
of $W_1/W_0$ \cite{aborny}, from the vortex free energy in a finite
volume \cite{KT1}, and from the adjoint string-breaking scale
\cite{dFP}.  The three determinations are in rough agreement.}  To
illustrate this fact, we generate two Gribov copies for each
thermalized lattice by applying the over-relaxation algorithm
both to the original thermalized lattice, and to a random gauge copy 
of that lattice.  The two gauge-fixed lattices are center projected
to obtain two
projected lattices, denoted $Z^A_\m(x)$ and $Z^B_\m(x)$.  We then
calculate projected Creutz ratios $\chi_{prod}(I,I)$ from the product
Wilson loops
\beq
      W_{prod}(C) = \langle Z^A(C) Z^B(C) \rangle
\eeq
If the two projected configurations were perfectly correlated, or 
if the $A$ and $B$ loops
differed only by perimeter effects, then we would find
\beq
       \chi_{prod}(I,I) = 0 ~~~~ \mbox{(strong correlation)}
\eeq
At the other extreme, if there were no correlation at all between
the projected configurations, we would find
\beq
       \chi_{prod}(I,I) = 2 \chi_A(I,I) = 2 \chi_B(I,I) 
                 ~~~~ \mbox{(no correlation)}
\eeq
where $\chi_A(I,I)=\chi_B(I,I)$ are the projected Creutz 
ratios obtained from $Z^A(C)$ or $Z^B(C)$ loops separately.

\FIGURE[h]{
\centerline{\scalebox{0.9}{
{\includegraphics{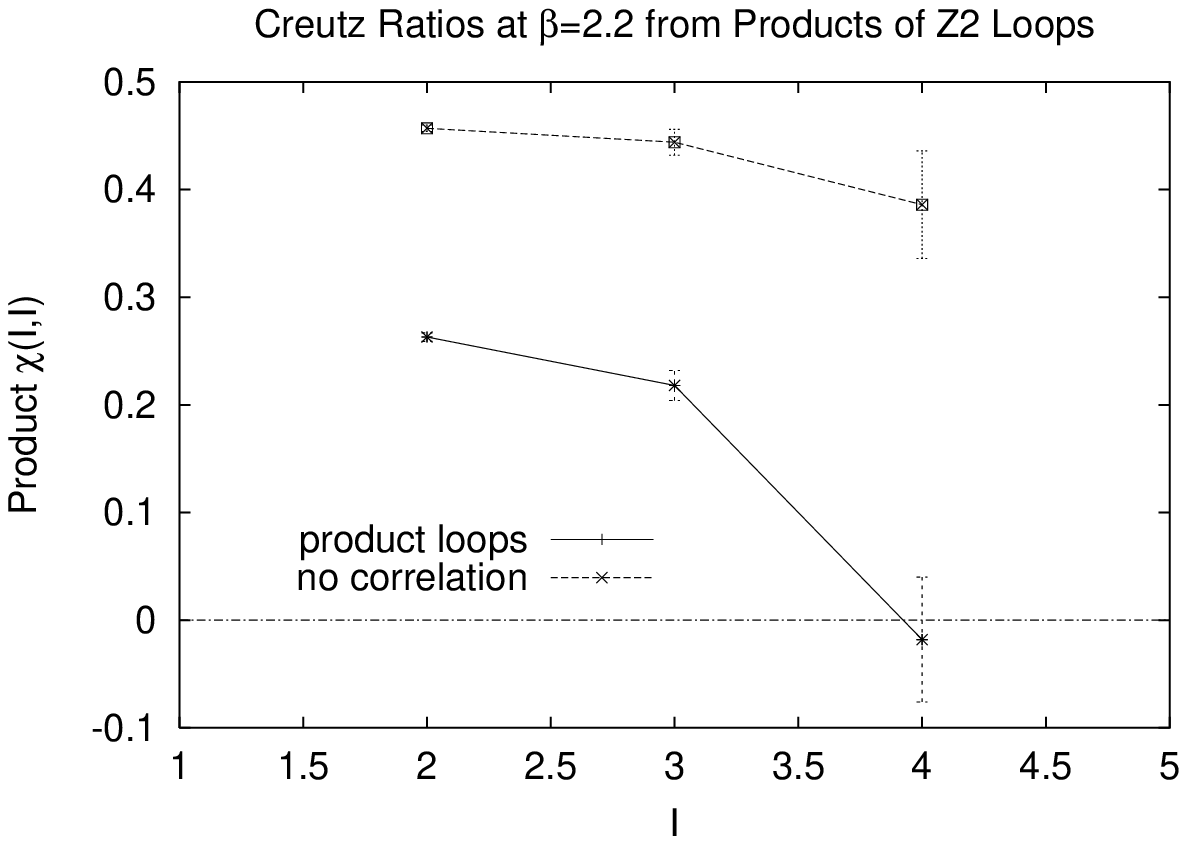}}}}
\caption{$\chi_{prod}(I,I)$ at $\b=2.2$.}
\label{corr22}
}

\FIGURE[h]{
\centerline{\scalebox{0.9}{
{\includegraphics{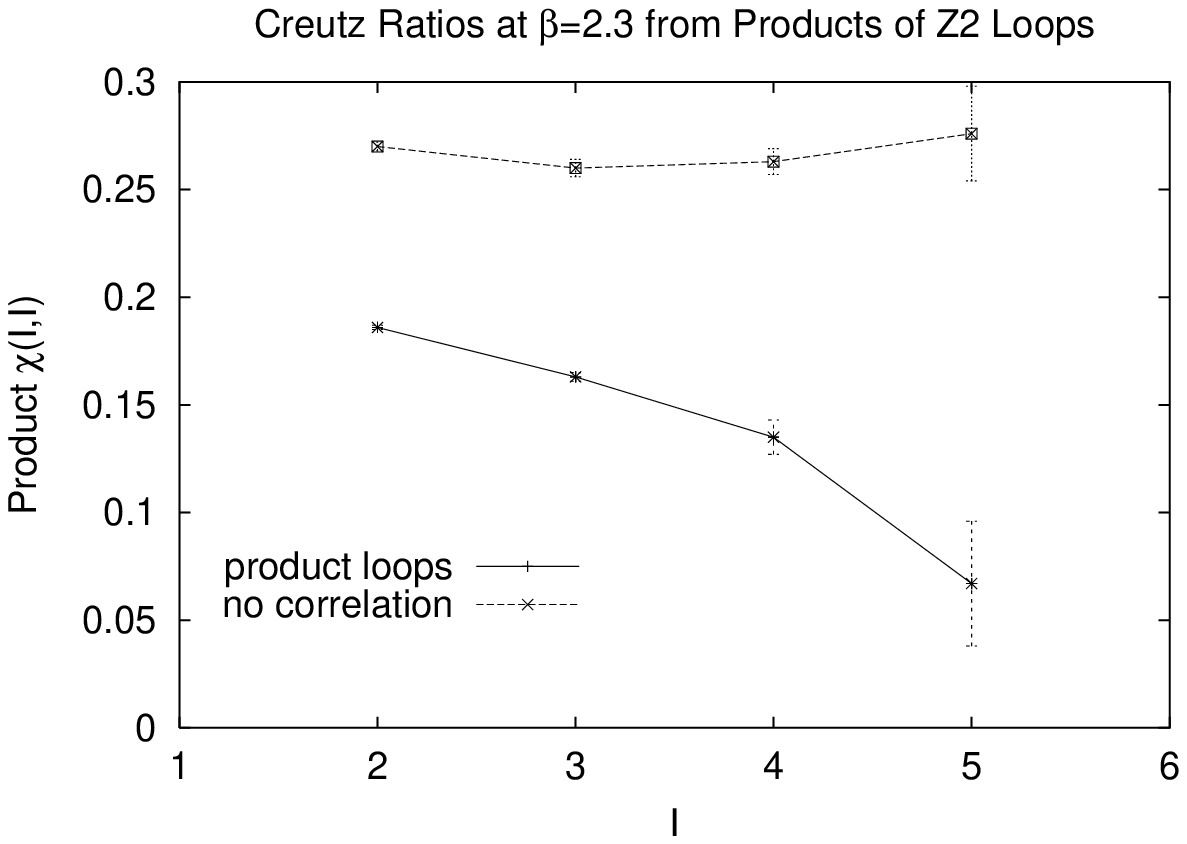}}}}
\caption{$\chi_{prod}(I,I)$ at $\b=2.3$.}
\label{corr23}
}

   Figures \ref{corr22} and \ref{corr23} 
show our results for $\chi_{prod}(I,I)$
at $\b=2.2$ and $\b=2.3$, which indicate a tendency towards
strong correlation of
the two projected Gribov copies in the infrared.

\section{Conclusions}         

   We have pointed out that there are at least two reasonable criteria for
selecting among Gribov copies in maximal center gauge.  One can select
the copy which is a best fit to a thin vortex over the entire lattice
volume, including links in P-vortices, and this leads to projected
string tensions which  
are some 34\% below the full string tension.
Alternatively, given that the fit is bound to fail at P-plaquettes,
one can select on the basis of the best fit in the lattice volume exterior
to P-plaquettes, and
in fact the exterior fit is better correlated with the projected string
tension.  But this second choice leads to projected string tensions which
are roughly 20\% higher ($\b=2.4$) than the full string tension.

     A case can be made for either criterion, but the method which we
have used in the past, now described as a process of ``quenched
minimization,'' seems to be something of a compromise between the two
alternatives: quenching emphasizes the external fit, subsequent
minimization the overall fit.  The projected string tensions obtained
in this way also come out roughly in the middle of the two extremes,
and have good center dominance properties.  Apart from center
dominance, the property of precocious linearity, and the scaling of
the vortex density, indicate that center projection has isolated the
relevant long-range degrees of freedom, rather than having them mixed
in with physics at all scales.  Projected lattices are also found to
be strongly correlated, in the infrared, among Gribov copies obtained
from the quenching procedure.

   Nevertheless, the variation in string tension among Gribov copies
selected according to different reasonable 
criteria is much greater than we had expected, and the justification
for quenched minimization is empirical rather than theoretical.
It is possible that an improved version of maximal center gauge can
be devised which retains the appealing ``best fit'' interpretation,
but which softens the contribution to the fitting functional at 
the location of P-vortices.  We consider this to be an interesting 
direction for further work.

\acknowledgments{

   Our research is supported in part by Fonds zur F\"orderung der
Wissenschaftlichen Forschung P13397-TPH (M.F.), the U.S. Department of 
Energy under Grant No.\ DE-FG03-92ER40711 (J.G.), the Slovak Grant 
Agency for Science, Grant No. 2/7119/2000 and the Slovak Literary
Fund (\v{S}.O.).  Our collaborative
effort is also supported by NATO Collaborative Linkage Grant No.\
PST.CLG.976987.  
}

\end{document}